\documentclass{article}
\begin{document}
\begin{center}
{\Large\bf {Lie Symmetries and Preliminary Classification  of
Group-Invariant Solutions  of Thomas equation}} {\vskip 0.5cm}
{\bf A. Ouhadan and E. H. El Kinani}\footnote{ Junior Associate
at The Abdus Salam ICTP, Trieste, Italy. e-mail hkinani@ictp.trieste.it}\\
{\vskip 0.5cm} Universit\'e Moulay Ismail, Facult\'e des Sciences
et Techniques, D\'epartement de Math\'ematiques, Laboratoire de
Physique
Math\'ematique, Boutalamine B.P.509, Errachidia, Morocco.\\

UFR de Sciences de l'Ing\'enieur, Facult\'e des Sciences et
Techniques, Boutalamine B.P.509, Errachidia, Morocco.\\
\end{center}
{\vskip 1cm}
\begin{abstract}
\hspace{.3in}  Using  the basic prolongation method and the
infinitesimal criterion of invariance, we find the most general
Lie point symmetries group of the Thomas equation. Looking the
adjoint representation of the obtained symmetry group on its Lie
algebra, we will find the preliminary classification of its
group-invariant solutions. This latter provides  a new exact
solutions for the Thomas equation.
\end{abstract}
{\vskip 1cm} \textbf {Keywords}: Lie-point symmetries, Thomas
equation, invariant
solutions.\\
\textbf {2000 Mathematics Subject Classification}:70G65, 58K70, 
34C14.

\newpage
\section {Introduction}
\hspace{.1in} The Lie group method is one of the most powerful
method available to analyse nonlinear evolution systems and their
methods of solution ${\cite{1, 2, 3, 4, 5}}$. Introduced firstly
by Sophus Lie ${\cite{6}}$ and consists of a systematic procedure
for the determination of continuous symmetry transformations of a
system of nonlinear partial differential equations (NLPDE). An
important feature of this method is that one can derive special
solutions associated with nonlinear PDEs straightforwardly which
are otherwise inaccessible through other methods. One of the
obvious advantages of knowing a symmetry group of a system of
differential equations is that one can use the defining propriety
of such a group and construct new solutions to the system from
known ones. One can also use symmetry group to explicitly
determine types of solution which are themselves invariant under
some subgroups of the full symmetry group of the system and
provides a new classification of different classes of solutions.
The basic idea of the Lie symmetry method is to find the
transformations groups associated with a given systems under a
continuous group of transformations and to find a reduction
transformation from the symmetries. In particular for nonlinear
partial differential equations with two independent variables the
reduction transformation can be used to reduce the number of
independent variables by one. In the case of the ordinary
differential equations the determination of a one-parameter
symmetry group allows  the reduction of the order of the equation
by one. Here using the criterion of invariance of the equation
under the infinitesimal prolonged generators, we find the most
general Lie point symmetries group of the Thomas equation. Looking
the adjoint representation of the obtained symmetry group on its
Lie algebra, we will find the preliminary classification of
group-invariant solutions. This  provides a new exact solutions of
the Thomas equation. The paper is arranged as follows. In section
2, we introduce the Thomas equation and by using the prolongation
formulae and the infinitesimal criterion of invariance we
determine the most general symmetry group and the corresponding
Lie algebra for Thomas equation. Section 3, is devoted to the
construction of the of group-invariant solutions and its
classification which provides in each case new exact solutions for
the Thomas equation. In section 4, we present  conclusions and
finally in the Appendix, we briefly give the methods for
constructing  the invariants of
the differential equation. \\

\section {Determination of Symmetry group for Thomas equation}
\subsection {Thomas equation}
\hspace{.1in} The Thomas equation is an interesting subject in
physical sciences ${\cite{7, 8, 9}}$it arises in the study of
chemical exchange processus and sometime called Thompson equation.
Proposed by Thomas ${\cite{9}}$ and has the form:

\begin{equation}
u_{xy}+\alpha u_{x}+\beta u_{y}+\gamma u_{x}u_{y}=0,
\end{equation}

where : $u_{x}=\frac{\partial u}{\partial x}, u_{y}=\frac{\partial
u}{\partial y}, u_{xy}=\frac{\partial^{2} u}{\partial x\partial
y}$ and $\alpha,\beta$ and $\gamma$ are constants. For exchange
processus $\alpha > 0,\beta>0$ and $\gamma\neq 0$.
\subsection {Lie algebra of Thomas equation}
\hspace{.1in}In order to calculate the Lie point symmetries of the
system(1), we first consider the generator of the group of point
transformations. Here there are two independent variables $x$ and
$y$,  and one dependent variable $u$. The Thomas equation is of
second order ($n=2$). Then a general vector field on the space of
independent and dependent variables ($X \times R$ where $X=<x,y>$)
takes the form :
\begin{equation}
v=\xi (x,y,u)\frac{\partial}{\partial x}+\eta
(x,y,u)\frac{\partial}{\partial y}+\varphi
(x,y,u)\frac{\partial}{\partial u}.
\end{equation}

Where  $\xi ,\eta$ and $\varphi$ depend on $x,y$ and  $u$ . We
wish to determine all possible coefficient functions $\xi ,\eta$
and $\varphi$, so that the corresponding one-parameter group $\exp
\epsilon v$ is a symmetry group of Thomas equation. According the
technics using in ${\cite{1}}$, since the Thomas equation is of
second order,  we need to know the second prolongation $Pr^{2}v$
of the infinitesimal generators which is given by the following
formulae :
\begin{equation}
Pr^{2}v=v+\varphi^{x}\frac{\partial}{\partial
u_{x}}+\varphi^{y}\frac{\partial} {\partial
u_{y}}+\varphi^{xx}\frac{\partial}{\partial
u_{xx}}+\varphi^{xy}\frac {\partial}{\partial
u_{xy}}+\varphi^{yy}\frac{\partial}{\partial u_{yy}},
\end{equation}
where the coefficients $\varphi^{x},\varphi^{y},\varphi^{xx},
\varphi^{xy}$ and $\varphi^{yy}$  are given by :

\begin{eqnarray}
\varphi^{x}&=&D_{x}\varphi -u_{x}D_{x}\xi -u_{y}D_{x}\eta, \\
\varphi^{y}&=&D_{y}\varphi -u_{x}D_{y}\xi -u_{y}D_{y}\eta, \\
\varphi^{xy}&=&D_{xy}(\varphi-\xi u_{x}-\eta u_{y})+\xi
u_{xxy}+\eta u_{xyy},\\
\varphi^{yy}&=&D_{yy}\varphi -2u_{xy}D_{y}\xi
-2u_{yy}D_{y}\eta-u_{x}D_{yy}\xi-u_{y}D_{yy}\eta, \\
\varphi^{xx}&=&D_{xx}\varphi -2u_{xx}D_{x}\xi
-2u_{xy}D_{x}\eta-u_{x}D_{xx}\xi-u_{y}D_{xx}\eta ,
\end{eqnarray}
where $D_{x}$ and  $ D_{y}$  are total derivatives on $x$ and $ y$
respectively and  $D_{xy}=D_{x}(D_{y})$. After a long calculation,
we obtain :
\begin{eqnarray}
\varphi^{x}&=&\varphi_{x}+(\varphi_{u} -\xi_{x})u_{x} -\eta_{x}
u_{y}-\xi_{u}u^{2}_{x}-\eta_{u} u_{x}u_{y}\\
\varphi^{y}&=&\varphi_{y}+(\varphi_{u} -\eta_{y})u_{y} -\xi_{y}
u_{x}-\eta_{u}u^{2}_{y}-\xi_{u} u_{x}u_{y} \\
\varphi^{xy}&=&\varphi_{xy}+(\varphi_{yu}-\xi_{xy})
u_{x}+(\varphi_{xu}-\eta_{xy})u_{y}-\xi_{yu}u_{x}^{2}\\
&+&(\varphi_{uu}-\xi_{ux}-\eta_{yu})
u_{x}u_{y}-\eta_{ux}u_{y}^{2}-\xi_{uu}u_{x}^{2}u_{y}-
\eta_{uu}u_{x}u_{y}^{2}\nonumber\\
&-&2\xi_{u}u_{x}u_{xy}-2\eta_{u}u_{y}u_{xy}-\xi_{y}u_{xx}-\eta_{x}
u_{yy}-\xi_{u}u_{y}u_{xx}-\eta_{u}u_{x}u_{yy}\nonumber\\
&+&(\varphi_{u}-\xi_{x}-\eta_{y}) u_{xy}\nonumber\\
\varphi^{xx}&=&\varphi_{xx}+(2\varphi_{xu}
-\xi_{xx})u_{x}-\eta_{xx}u_{y}-2\eta_{xu}u_{x}u_{y}\\\nonumber
&+&(\varphi_{uu}-2\xi_{xu})u_{x}^{2}-\eta_{uu}u_{x}^{2}u_{y}-2\eta_{x}u_{xy}\\\nonumber
&-&2\eta_{u}u_{x}u_{xy}+(\varphi_{u}
-2\xi_{x})u_{xx}-3\xi_{u}u_{x}u_{xx}-\xi_{uu}u_{x}^{3}\\\nonumber
&-&\eta_{u}u_{y}u_{xx}\nonumber\\
\varphi^{yy}&=&\varphi_{yy}+(2\varphi_{yu}
-\eta_{yy})u_{y}-\xi_{yy}u_{x}-2\xi_{yu}u_{x}u_{y}\\\nonumber
&+&(\varphi_{uu}-2\eta_{yu})u_{y}^{2}-\xi_{uu}u_{y}^{2}u_{x}-2\xi_{y}u_{xy}\\\nonumber
&-&2\xi_{u}u_{y}u_{xy}+(\varphi_{u}
-2\eta_{y})u_{yy}-3\eta_{u}u_{y}u_{yy}-\eta_{uu}u_{y}^{3}\\\nonumber
&-&\xi_{u}u_{x}u_{yy}\nonumber
\end{eqnarray}
To use the infinitesimal criterion of invariance, let us introduce
$\Delta=u_{xy}+\alpha u_{x}+\beta u_{y}+\gamma u_{x}u_{y}$.\\
Suppose that :
\begin{equation}
\Delta=0, \hspace{.2in} \mbox{if } \hspace{.1in}
Pr^{2}v(\Delta)=0,
\end{equation}
for every infinitesimal generators $v$ of $G$, then $G$ is a
symmetry group of the Thomas equation (1). Substituting the
general formulae (9, 10 ) and (11)  into (3) and (14), after
eliminating the dependence between the derivatives coming from the
equation itself and equating the coefficients of the various
monomials in the first and second order, we find the defining
equations for the symmetry group of the Thomas equation to be the
following :
\begin{center}
\begin{tabular}{| r| l|}
\hline
Monomial & coefficient \\
\hline
1 & $\alpha \varphi_{x}+\beta \varphi_{y}+\varphi_{xy}$\\
\hline
$u_{x}$&$\varphi_{yu}-\xi_{xy}+\gamma\varphi_{y}-\beta\xi_{y}+\alpha\eta_{y}$\\
\hline
$u_{y}$&$\varphi_{xu}-\eta_{xy}+\gamma\varphi_{x}-\alpha\eta_{x}+\beta\xi_{x}$\\
\hline
$u_{x}u_{y}$&$\varphi_{uu}-\xi_{ux}-\eta_{yu}+\beta\xi_{u}+\alpha\eta_{u}+\gamma\varphi_{u}$\\
\hline
$u_{x}^{2}$&$-\xi_{yu}-\gamma\xi_{y}+\alpha\xi_{u}$\\
\hline
$u_{y}^{2}$&$-\eta_{xu}-\gamma\eta_{x}+\beta\eta_{u}$\\
\hline
$u_{x}^{2}u_{y}$&$-\xi_{uu}$\\
\hline
$u_{x}u_{y}^{2}$&$-\eta_{uu}$\\
\hline
$u_{xx}$&$-\xi_{y}$\\
\hline
$u_{yy}$&$-\eta_{x}$\\
\hline
$u_{y}u_{xx}$&$-\xi_{u}$\\
\hline
$u_{x}u_{yy}$&$-\eta_{u}$\\
\hline
\end{tabular}
\end{center}
\begin {center}{\it{Table 1: Coefficients of various monomials}}
\end{center}

The system  determining the  symmetry group of our equation is :\\

\begin{eqnarray}
\alpha \varphi_{x}+\beta \varphi_{y}+\varphi_{xy}&=&0;
\hspace{3cm}(15-1)\nonumber\\
\varphi_{yu}-\xi_{xy}+\gamma\varphi_{y}-\beta\xi_{y}+\alpha\eta_{y}&=&0;\hspace{3cm}(15-2)\nonumber\\
\varphi_{xu}-\eta_{xy}+\gamma\varphi_{x}-\alpha\eta_{x}+\beta\xi_{x}&=&0;\hspace{3cm}(15-3)\nonumber\\
\varphi_{uu}-\xi_{ux}-\eta_{yu}+\beta\xi_{u}+\alpha\eta_{u}+\gamma\varphi_{u}&=&0;\hspace{3cm}(15-4)\nonumber\\
-\xi_{yu}-\gamma\xi_{y}+\alpha\xi_{u}&=&0;\hspace{3cm}(15-5)\nonumber\\
-\eta_{xu}-\gamma\eta_{x}+\beta\eta_{u}&=&0;\hspace{3cm}(15-6)\nonumber\\
-\xi_{uu}&=&0;\hspace{3cm}(15-7)\nonumber\\
-\eta_{uu}&=&0;\hspace{3cm}(15-8)\nonumber\\
-\xi_{y}&=&0;\hspace{3cm}(15-9)\nonumber\\
-\eta_{x}&=&0;\hspace{3cm}(15-10)\nonumber\\
-\xi_{u}&=&0;\hspace{3cm}(15-11)\nonumber\\
-\eta_{u}&=&0;\hspace{3cm}(15-12)\nonumber
\end{eqnarray}
From (15-9) to (15-12) we obtain : $$\xi=\xi(x);\textrm{ \quad and
\quad
}\eta=\eta(y)$$ \\
From (15-4) we deduce that : $$\varphi(x,y,u)=g(x,y)\exp^{-\gamma
u}+N(x,y)$$\\ For convenient we can chose $\frac{-g(x,y)}{\gamma}$
for
$g(x,y)$.\\
Using the above results and (15-2) with (15-3) one finds :
$$\gamma
N_{x}+\beta\xi_{x}=0;\:\:\:\:\gamma N_{y}+\alpha\eta_{y}=0,$$\\
which yields to : $$N_{xy}=0; \quad  \xi_{x}=-\frac{\gamma
N_{x}}{\beta}, \textrm{ and }\eta_{y}=-\frac{\gamma
N_{y}}{\alpha},$$ so we have \quad $N(x,y)=mx+ny+a  $ \quad  where
$m, n $ and $ a$  are arbitrary constants, the linear first
partial differential equation
(15-1) leads to \quad $\alpha m+\beta n=0.$\\
Then, for \quad \quad $k=\frac{m}{\beta}, $\quad \quad one
obtains the most general form of the coefficient functions $\xi,
\eta $ and $ \varphi$
which are the form :\\
\begin{eqnarray}
\xi(x,y,u)&=&-k\gamma x+c,\\
\eta(x,y,u)&=&k\gamma y+b,\\
\varphi(x,y,u)&=&-\frac{g(x,y)}{\gamma}e^{-\gamma u}+k(\beta
x-\alpha y)+a,
\end{eqnarray}
where $a, b, c$ and $k$  are arbitrary constants and $g(x,y)$
an arbitrary solution of the  equation  (15-1).
Then, we obtain the most general form
for vector field $v$ which is :
\begin{equation}
v=(-k\gamma x+c)\frac{\partial}{\partial x}+(k\gamma
y+b)\frac{\partial}{\partial y}+ (-\frac{g(x,y)}{\gamma}e^{-\gamma
u}+k(\beta x-\alpha y)+a)\frac{\partial}{\partial u}.
\end{equation} Hence, the Lie algebra of infinitesimal symmetry of the
Thomas
equation is
spanned by the four vectors fields :\\
\begin{eqnarray}
v_{1}&=&\frac{\partial}{\partial x};\\
v_{2}&=&\frac{\partial}{\partial y};\\
v_{3}&=&\frac{\partial}{\partial u};\\
v_{4}&=&-\gamma x\frac{\partial}{\partial x}+\gamma
y\frac{\partial}{\partial y}+ (\beta x-\alpha y)\frac{\partial}{\partial
u};\nonumber\\
\end{eqnarray}
and the infinite dimensional subalgebra :
\begin{equation}
v_{g}=-\frac{g(x,y)}{\gamma}e^{-\gamma u}\frac{\partial}{\partial
u}.
\end{equation}

The one-parameter groups $G_{i}$ generated by the $v_{i}$ are
given in the following table. The entries gives the transformed
point $\exp(\epsilon
v_{i})(x,y,u)=(\widetilde{x},\widetilde{y},\widetilde{u})$ :
\begin{eqnarray}
G_{1} &:& (x+\epsilon,y,u)\\
G_{2} &:& (x,y+\epsilon,u)\\
G_{3} &:& (x,y,u+\epsilon)\\
G_{4} &:&
(xe^{-\gamma\epsilon},ye^{\gamma\epsilon},\frac{\beta}{\gamma}
x(1-e^{-\gamma\epsilon})+\frac{\alpha}{\gamma}y(1-e^{\gamma\epsilon})+u)\\
G_{g}&:&(x,y,\frac{1}{\gamma}\log[\gamma g(x,y)\epsilon+e^{\gamma
u }])
\end{eqnarray}
Since each group $G_{i}$ is a symmetry group, then if $u=f(x,y)$
is a solution of the Thomas equation, so are the functions :
\begin{eqnarray}
u_{1}&=&f(x-\epsilon,y)\\
u_{2}&=&f(x,y-\epsilon,)\\
u_{3}&=&f(x,y)+\epsilon\\
u_{4}&=&\frac{\beta}{\gamma}x(e^{\gamma\epsilon}-1)+\frac{\alpha}{\gamma}
y(e^{-\gamma\epsilon}-1)+f(xe^{\gamma\epsilon},ye^{-\gamma\epsilon})\\
u_{g}&=&\frac{1}{\gamma}\log[\gamma g(x,y)\epsilon+ e^{\gamma
f(x,y)}].
\end{eqnarray}
where $\epsilon $ is any number real.\\ The commutation relations
between these vector fields are given in the following table :
\begin{center}
\begin{tabular}{|r|r|c|l|l|l|}
\hline
$[v_{i},v_{j}]$&$v_{1}$&$v_{2}$&$v_{3}$&$ v_{4}$&$v_{f}$\\
\hline
$v_{1}$&0&0&0&$-\gamma v_{1}+\beta v_{3}$&$v_{f_{x}}$\\
\hline
$v_{2}$&0&0&0&$\gamma v_{2}-\alpha v_{3}$&$v_{f_{y}}$\\
\hline
$v_{3}$&0&0&0&0&$v_{-\gamma f}$\\
\hline $v_{4}$&$\gamma v_{1}-\beta v_{3}$&$-\gamma v_{2}+\alpha
v_{3}$&0&0&$v_{\psi}$\\
\hline $v_{g}$&$-v_{f_{x}}$&$-v_{f_{y}}$&$v_{\gamma
f}$&$v_{\psi}$&0\\
\hline
\end{tabular}
\end{center}
\begin {center}{\it{Table 2: Commutator table for the Lie algebra $v_{i}$
and $v_{g}$ }}
\end{center}
where $$ \psi=-\gamma
xf_{x}+\gamma yf_{y}-\gamma(\beta x-\alpha y)f.$$\\
Since the totality of infinitesimal symmetries must be
a Lie algebra, we can conclude that if $f(x,y)$ is any solution of
the  equation in (15-1), so are $f_{x}$, $f_{y}$ and $-\gamma
xf_{x}+\gamma yf_{y}-\gamma(\beta x-\alpha y)f$.

\section {Classification of  Group-Invariant Solutions}
\hspace{.1in}Recall first that in general  to each one parameter
subgroups of the full symmetry group of a system there will
corresponding a family of solutions,  such solutions are called
invariant solutions ${\cite{1}}$. In this paper we are interesting
only on the symmetry algebra $g$ of the Thomas equation which is
spanned by the vector fields $v_{1},v_{2},v_{3}$ and $v_{4}$.
Every one-dimentional subalgebra of  $g$ is determined by a
nonzero vector $v$ of the form :
\begin{equation}
v=a_{1}v_{1}+a_{2}v_{2}+a_{3}v_{3}+a_{4}v_{4},
\end{equation}
where $a_{i}$ are arbitrary constants. Our task is to simplify as
many of the coefficients $a_{i}$ as possible through application
of adjoint to $v$. To compute the adjoint representation, we use
the Lie  series :
\begin{equation}
Ad(exp(\epsilon
v_{i}))v_{j}=v_{j}-\epsilon[v_{i},v_{j}]+\frac{\epsilon^{2}}{2}[v_{i},[v_{i},v_{j}]]-\cdots
\end{equation}
Then from the commutation table 2, we obtain the following table :
\begin{center}
\begin{tabular}{|r|r|c|l|l|}
\hline
$Ad$&$v_{1}$&$v_{2}$&$v_{3}$&$ v_{4}$\\
\hline
$v_{1}$&$v_{1}$&$v_{2}$&$v_{3}$&$v_{4}+\epsilon\gamma v_{1}-\epsilon\beta
v_{3}$\\
\hline
$v_{2}$&$v_{1}$&$v_{2}$&$v_{3}$&$v_{4}-\epsilon\gamma v_{2}+\epsilon\alpha
v_{3}$\\
\hline
$v_{3}$&$v_{1}$&$v_{2}$&$v_{3}$&$v_{4}$\\
\hline $v_{4}$&$e^{-\gamma\epsilon}
v_{1}+\frac{\beta}{\gamma}(1-e^{-\gamma\epsilon})
v_{3}$&$e^{\gamma\epsilon}
v_{2}-\frac{\alpha}{\gamma}(e^{-\gamma\epsilon}-1)v_{3}$
&$v_{3}$&$v_{4}$\\
\hline
\end{tabular}
\end{center}
\begin {center}{\it {Table 3: Adjoint table for the Lie algebra $v_{i}$}}
\end{center}

In what follows, we begin the classification process :\\

\noindent{}\underline{\textbf{Case.1}}\\

Suppose first that  $a_{4}\ne 0$, we can assume that  $a_{4}=1$,
and the vector field
$v$ takes the form :\\
\begin{equation}
v=a_{1}v_{1}+a_{2}v_{2}+a_{3}v_{3}+v_{4}
\end{equation}

Refereing to the table.3, if we act on such a $v$ by
$Ad(exp(\frac{a_{3}}{\beta})v)$, then we can make the coefficient
$v_{3}$ vanish . Then the reduced vector field takes the form :

\begin{equation}
v=(a_{1}-\gamma x)\frac{\partial}{\partial x}+(a_{2}+\gamma y)
\frac{\partial}{\partial y}+(\beta x-\alpha
y)\frac{\partial}{\partial u}.
\end{equation}
The invariants $\varsigma$ and  $\chi$  can be found by
integrating the corresponding characteristic system (s\emph{ee Appendix
for construction of the invariants}), which is  :\\
\begin{equation}
\frac{dx}{a_{1}-\gamma x}=\frac{dy}{a_{2}+\gamma y}=
\frac{du}{\beta x-\alpha y}
\end{equation}
The obtained solution are given by
\begin{equation}
\chi=(a_{1}-\gamma x)(a_{2}+\gamma y),
\end{equation}
and
\begin{equation}
\varsigma =u+\frac{\beta}{\gamma}x+\frac{(\beta a_{1}+\alpha
a_{2})}{\gamma^{2}}log(\gamma x-a_{1})-\frac{\alpha
\chi}{\gamma^{2}(\gamma x-a_{1})}.
\end{equation}

Therefore,  a solution of our equation in this case is :
\begin{equation}
u=f(x,\chi,\varsigma)=\varsigma-(\frac{\beta a_{1}+\alpha
a_{2}}{\gamma^{2}})log(\gamma x-a_{1})+\frac{\alpha\chi}
{\gamma^{2}(\gamma x-a_{1})^{2}}-\frac{\beta}{\gamma}x
\end{equation}
The derivatives of $u$ are given in terms of $\varsigma$ and $\chi$ as :\\
\begin{eqnarray}
u_{x}&=&-\gamma(a_{2}+\gamma y)\varsigma _{\chi}-\frac{\beta
a_{1}+\alpha a_{2}}{\gamma(\gamma x-a_{1})}-
\frac{\beta}{\gamma};\\
u_{y}&=&-\frac{\alpha}{\gamma}+\gamma(a_{1}-\gamma x)\varsigma _{\chi};\\
u_{xy}&=&-\gamma^{2}\varsigma _{\chi}-\gamma^{2} \varsigma
_{\chi\chi};
\end{eqnarray}

Substituting (42), (43) and (44) into  the Thomas equation we obtain
reduced equation  of  the form :\\
\begin{equation}
\gamma^{2}\chi \varsigma_{\chi\chi}+\gamma^{3}\chi
\varsigma_{\chi}^{2}+\gamma^{2}(\gamma-\beta a_{1}- \alpha
a_{2})\varsigma_{\chi}+\frac{\alpha\beta}{\gamma}=0.
\end{equation}
For $\varsigma_{\chi}=\theta$, one obtains the following
equation :
\begin{equation}
\theta'=-\gamma\theta^{2}-\frac{(\gamma-\beta a_{1}-\alpha
b_{1})}{\gamma}\frac{\theta}
{\chi}+\frac{\alpha\beta}{\gamma^{2}\chi}=0;
\end{equation}

\noindent{this} Riccati type equation is transformed by using the
change :
\begin{equation}
y(\chi)=\exp(\int \gamma \theta(\chi)),
\end{equation}
 multiplying by $\chi$ one obtain  the Fuchs type equation :
\begin{equation}
\chi^{2} y^{''}+e\chi y^{'}+m\chi y=0,
\end{equation}

\noindent{where}  $e=\frac{(\gamma-\beta a_{1}-\alpha
a_{2})}{\gamma}$
and  $m=\frac{\alpha\beta}{\gamma^{2}}$.\\

This equation admits a solution developable in series in the
neighboring of zero,  for $\chi > 0$, we have :
\begin{equation}
y_{p}(\chi)=\sum_{n=0}^{\infty}a_{n}\chi^{n},
\end{equation}

where   $a_{0}=1$ and  $a_{n}=\frac{(-m)^{n}}{n!e(e+1)\cdots(e+n-1)}$, $n\ge
1$\\

In the case  $\chi < 0$ by introducing the transformation   $\chi
\longmapsto -\chi$, This series have an infinite radius of
convergence for all  $\chi > 0$, then
if we introduce the change $z_{p}(\chi)=\frac{y_{p}^{'}}{\gamma y_{p}}$,
$z_{p}$
is a particular solution of the equation (46).
Putting  $\theta=z+z_{p}$, (46) becomes a Bernoulli type equation
:
\begin{equation}
z^{'}=-\gamma z^{2}-2\gamma z_{p}z,
\end{equation}
which is transformed in the linear equation (with $z=\frac{1}{f}$)
\begin{equation}
f^{'}=2\gamma z_{p}f+\gamma.
\end{equation}
The solution of (51) is :
\begin{equation}
f(\chi)=g_{p}(\chi)y_{p}^{2},
\end{equation}
where $g_{p}$ ia a primitive of $\frac{1}{\gamma y_{p}^{2}}$\\
Therefore,
\begin{equation}
\theta(\chi)=\frac{1}{f(\chi)}+z_{p}(\chi).
\end{equation}
Thus, we have $\varsigma=\varsigma(\chi)=\int\theta(\chi)d\chi$, \\
and then, we obtain the solution in this case which is :
\begin{equation}
u=\varsigma(\chi)-(\frac{\beta a_{1}+\alpha
a_{2}}{\gamma^{2}})log(\gamma
x-a_{1})+\frac{\alpha\chi}{\gamma^{2}(\gamma
x-a_{1})}-\frac{\beta}{\gamma}x,
\end{equation}
where $\chi=(a_{1}-\gamma x)(a_{2}+\gamma y)$.\\
\newpage

\noindent{}\underline{\textbf{Case.2}}\\

Suppose now that  $a_{4}=0$, and  $a_{3}\ne 0$. In this situation
the
invariants associated to the field vector $v$ are of the form :\\

\noindent{}\underline{\it{Case 2.1}} : If $a_{1}a_{2}\ne 0,$
\begin{eqnarray}
v&=&a_{1}v_{1}+a_{2}v_{2}+v_{3};\\
&=&a_{1}\frac{\partial}{\partial x}+a_{2}\frac{\partial}{\partial
y}+\frac{\partial}{\partial u}\nonumber;
\end{eqnarray}
The invariants are :
\begin{eqnarray}
\chi&=&a_{2}x-a_{1}y;\\
\varsigma&=&u-\frac{y}{a_{2}};
\end{eqnarray}
Next, we have :
\begin{eqnarray}
u_{x}&=&a_{2}\varsigma_{\chi};\\
u_{y}&=&\frac{1}{a_{2}}-a_{1}\varsigma_{\chi};\\
u_{xy}&=&-a_{1}a_{2} \varsigma_{\chi\chi}.
\end{eqnarray}
Substituting these quantities  in the Thomas equation, we find  the
reduced equation :
\begin{equation}
-a_{1}a_{2}\varsigma_{\chi\chi}+(\alpha a_{2}-\beta
a_{1}+\gamma)\varsigma_{\chi}-a_{1}a_{2}\gamma
\varsigma_{\chi}^{2}+\frac{\beta}{a_{2}}=0;
\end{equation}
Putting  $\varsigma_{\chi}=\theta$, then (61) takes  the form :
\begin{equation}
-a_{1}a_{2}\theta'+(\alpha a_{2}-\beta
a_{1}+\gamma)\theta-a_{1}a_{2}\gamma\theta^{2}+\frac{\beta}{a_{2}}=0;
\end{equation}
which is the Riccati type equation.\\

\noindent{}\underline{\it{Case 2.1.a}}:  If  $(\alpha a_{1}-\beta
a_{1}+\gamma)^{2}+4a_{1}\gamma\beta \geq 0$.\\

Then, our equation admit one constant solution $\theta_{0}$ which
satisfying the second order equation:
\begin{equation} (\alpha a_{1}-\beta
a_{1}+\gamma)z-a_{1}a_{2}\gamma z^{2}+\frac{\beta}{a_{2}}=0.
\end{equation}
Then by redefining the variable $\theta=f+\theta_{0}$, one obtain
the following Bernoulli type equation  :
\begin{equation}
-a_{1}a_{2}f'+(\alpha a_{2}-\beta
a_{1}+\gamma-2a_{1}a_{2}\gamma\theta_{0})f-a_{1}a_{2}\gamma
f^{2}=0;
\end{equation}
Put now $f=\frac{1}{h}$ one obtain a linear equation:
\begin{equation}
h'=\frac{2a_{1}a_{2}\gamma\theta_{0}-\alpha a_{2}+\beta a_{1}
+\gamma}{a_{1}a_{2}}h+\gamma
\end{equation}
This equation admit a solution, since
$\frac{2a_{1}a_{2}\gamma\theta_{0}-\alpha a_{2}+\beta a_{1}
+\gamma}{a_{1}a_{2}}\ne 0$, which is :
\begin{equation}
h=Aexp(\frac{2a_{1}a_{2}\gamma\theta_{0}-\alpha a_{2}-\beta a_{1}
+\gamma}{a_{1}a_{2}}\chi)- \frac{\gamma
a_{1}a_{2}}{2a_{1}a_{2}\gamma\theta_{0}-\alpha a_{1}-\beta a_{1}
+\gamma}
\end{equation}
where  $A$ is an arbitrary constant. Hence,
\begin{equation}
\varsigma=\int\frac{1}{Aexp(\frac{2a_{1}a_{2}\gamma\theta_{0}-\alpha
a_{2}+\beta a_{1} +\gamma} {a_{1}a_{2}}\chi)-\frac{\gamma
a_{1}a_{2}}{2a_{1}a_{2}\gamma\theta_{0}-\alpha a_{1}-\beta a_{1}
+\gamma}}d\chi+\theta_{0}\chi+cte
\end{equation}
After integration we obtain :
\begin{equation}
\varsigma=\frac{1}{\gamma}log(A-\frac{\gamma
}{C}e^{-C\chi})+\theta_{0}\chi+cte
\end{equation}
where  $C=\frac{2a_{1}a_{2}\gamma\theta_{0}-\alpha a_{2}+\beta
a_{1}-\gamma}{a_{1}a_{2}} $. Finally, the solution of the  Thomas
equation is:
\begin{equation}
u=\frac{1}{\gamma}log(A-\frac{\gamma}{C}
e^{-C(a_{2}x-a_{1}y)}+\theta_{0}(a_{2}x-a_{1}y)+\frac{y}{a_{2}}+cte;
\end{equation}
\\

\noindent{}\underline{\it{Case 2.1.b}}: If  $(\alpha a_{2}-\beta
a_{1}+\gamma)^{2}+4a_{1}\gamma\beta < 0$.\\

\noindent{In} this situation we have :
\begin{equation}
\frac{(\alpha a_{2}-\beta
a_{1}+\gamma)}{a_{1}a_{2}}\theta-\gamma\theta^{2}+\frac{\beta}{a_{1}a_{2}^{2}}
\ne 0.
\end{equation}
The reduced equation obtained can be written in the separation
form as :
\begin{equation}
\frac{d\theta}{\frac{(\alpha a_{2}-\beta
a_{1}+\gamma)}{a_{1}a_{2}}\theta-\gamma\theta^{2}+\frac{\beta}{a_{1}a_{2}^{2}}}=dx.
\end{equation}
If we put   $A_{1}=\frac{(\alpha a_{2}-\beta
a_{1}+\gamma)}{a_{1}a_{2}}, A_{2}=-\gamma$, and  $A_{3}=
\frac{\beta}{a_{1}a_{2}^{2}}$, thus the problem becomes to the
integration of the following equation :
\begin{equation}
\frac{d\theta}{A_{1}\theta+A_{2}\theta^{2}+A_{3}}=dx.
\end{equation}
Let   $\Xi = \frac{4A_{2}A_{3}-A{1}^{2}}{4A_{2}^{2}} $, $\Xi
>0$. Then, we look to the integration of
\begin{equation}
\frac{1}{(\theta +\frac{A_{1}}{2A_{2}})^{2}+\Xi}.
\end{equation}
By using the change of the  variable $t=\Xi
+\frac{A_{1}}{2A_{2}}$, we have :
\begin{equation}
\int \frac{d\theta}{(\theta
+\frac{A_{1}}{2A_{2}})^{2}+\Xi}=\frac{1}
{\sqrt{\Xi}}arctang(\frac{\theta+\frac{A_{1}}{2A_{2}}}{\sqrt{\Xi}})+cte;
\end{equation}
and therefore,
\begin{equation}
\frac{1}{\sqrt{\Xi}}arctang(\frac{\theta+\frac{A_{1}}{2A_{2}}}{\sqrt{\Xi}})=A_{2}x+cte,
\end{equation}
then,
\begin{equation}
\theta=\sqrt{\Xi}tg(A_{2}\sqrt{\Xi}x+cte)-\frac{A_{1}}{2A_{2}},
\end{equation}
and we obtain :
\begin{eqnarray}
\varsigma_{\chi}&=&\sqrt{\Xi}tg(A_{2}\sqrt{\Xi}\chi+cte)-\frac{A_{1}}{2A_{2}},\\
\varsigma&=&\sqrt{\Xi}\int
tg(A_{2}\sqrt{\Xi}\chi+cte)d\chi-\frac{A_{1}}{2A_{2}}\chi+cte, \\
&=&\frac{1}{2A_{2}}log(1+tg(A_{2}\sqrt{\Xi}\chi+cte)-\frac{A_{1}}{2A_{2}}\chi+cte.\nonumber
\end{eqnarray}
Next, we find the solution of the Thomas equation which is :

\begin{equation}
u=\frac{1}{2A_{1}}log(1+tg(A_{2}\sqrt{\Xi}(a_{2}x-a_{1}y)+A_{0}
)-\frac{A_{1}}{2A_{2}}(a_{2}x-a_{1}y)+\frac{y}{a_{2}}+cte;
\end{equation}
where $A_{0}=$ constant.\\

\noindent{}\underline{\it{Case 2.2}} : If $a_{2}=0, a_{1}\ne 0$ and  $a_{1}
\ne \frac{-\gamma}{\beta}$\\

\noindent{In} this case the vectors field is:
\begin{equation}
v=a_{1}\frac{\partial}{\partial x}+\frac{\partial}{\partial u}.
\end{equation}
The  invariants are :
\begin{eqnarray}
\chi &=&y,\\
\varsigma &=&-a_{1}u+x,
\end{eqnarray}
then, $u=\frac{-\chi}{a_{1}}+\frac{x}{a_{1}}$, and :
\begin{eqnarray}
u_{x}&=&\frac{1}{a_{1}},\\
u_{y}&=&\frac{-\varsigma_{\chi}}{a_{1}},\\
u_{xy}&=&0.
\end{eqnarray}
After substituting these expressions into the Thomas equation, we
obtain the following reduced equation :
\begin{equation}
(\beta a_{1}+\gamma)\varsigma_{\chi}=a_{1}\alpha;
\end{equation}
and then,
\begin{equation} \varsigma=\frac{a_{1}\alpha}{\beta
a_{1}+\gamma}\chi +cte.
\end{equation}
Next we have,
\begin{equation}
u=\frac{x}{a_{1}}-\frac{\alpha}{\beta a_{1}+\gamma}\varsigma+cte;
\end{equation}
and therefore,
\begin{equation}
u=\frac{x}{a_{1}}-\frac{\alpha}{\beta a_{1}+\gamma}y+cte.
\end{equation}
\\

\noindent{}\underline{\it{Case 2.3}} : If
$a_{1}=\frac{-\gamma}{\beta}$.\\

 \noindent{The} vector field $v$ take the
form :
\begin{equation}
v=\frac{-\gamma}{\beta}\frac{\partial}{\partial
x}+\frac{\partial}{\partial u}.
\end{equation}
and the invariants are :
\begin{eqnarray}
\chi&=&y;\\
\varsigma&=&\beta x+\gamma u.
\end{eqnarray}
Therefore
\begin{equation}
u=\frac{\chi}{\gamma}-\frac{\beta}{\gamma}x,
\end{equation} and then,
\begin{eqnarray}
u_{x}&=&\frac{-\beta}{\gamma};\\
u_{y}&=&\frac{-\varsigma_{\chi}}{\gamma};\\
u_{xy}&=&0.
\end{eqnarray}
Substituting the above expressions into Thomas equation, one find
that  $\alpha\beta=0$. Then we conclude that there is not the
solution in this case.\\

\noindent{}\underline{\it{Case 2.4}} : If $a_{1}= a_{2}=0$.\\

\noindent{we} have :
\begin{equation}
v=\frac{\partial}{\partial u};
\end{equation}
the  invariants are $\varsigma= x$, and $\chi = y$, then like in
the last case there is not the solution of the equation.\\

\noindent{}\underline{\textbf{Case.3}}\\

In this case we choose  $a_{4}=a_{3}=0,$ then the vector field is
the form :
\begin{equation}
v=a_{1}\frac{\partial}{\partial x}+a_{2}\frac{\partial}{\partial
y};
\end{equation}
\\

\noindent{}\underline{\it{Case 3.1}} :  $a_{1}a_{2} \ne 0$.\\

\noindent{The} vector field is of the form :
\begin{equation}
v=\frac{\partial}{\partial x}+a_{2}\frac{\partial}{\partial y};
\end{equation}
and the  invariants $\varsigma$ et $\chi$ are solutions of the
following characteristics system  :
\begin{equation}
\frac{dx}{1}=\frac{dy}{a_{2}}=\frac{du}{0}.
\end{equation}
The solutions are :
\begin{eqnarray}
\chi&=&x-\frac{y}{a_{2}};\\
\varsigma&=&u.
\end{eqnarray}
next we have : $u=\varsigma$, therefore :
\begin{eqnarray}
u_{x}&=& \varsigma_{\chi};\\
u_{y}&=&\frac{-\varsigma_{\chi}}{a_{2}};\\
u_{xy}&=&\frac{-\varsigma_{\chi\chi}}{a_{2}};
\end{eqnarray}
By substituting these  expressions into the Thomas equation, we
obtain the following equation :
\begin{equation}
\varsigma_{\chi\chi}+(\beta-a_{2}\alpha)\varsigma_{\chi}+\gamma
\varsigma_{\chi}^{2}=0;
\end{equation}
putting  $\theta = \varsigma_{\chi}$, we obtain the Bernoulli
equation :
\begin{equation}
\theta'+(\beta-a_{2}\alpha)\theta+\gamma\theta^{2}=0
\end{equation}
\newpage

\noindent{}\underline{\it{Case 3.1.a}} : If $a_{2} =
\frac{\beta}{\alpha}$.\\

\noindent{The} reduced equation takes the form :

\begin{equation}
\theta'+\gamma\theta^{2}=0;
\end{equation}
which have the solution :
\begin{equation}
\theta =\frac{1}{\gamma\chi+cte},
\end{equation}
and therefore,
\begin{equation}
u=\frac{1}{\gamma}log(\gamma(x-\frac{y}{a_{2}})+k_{0})+cte;
\end{equation}
where $k_{0}$, is arbitrary constant.\\

\noindent{} \underline{\it{Case 3.1.b}} : If $a_{2} \ne
\frac{\beta}{\alpha}$.\\

\noindent{We} adopt the change $\theta=\frac{1}{z}$, we obtain the
following linear equation :
\begin{equation}
z'=(\beta-a_{2}\alpha)z+\gamma,
\end{equation}
which admit the general solution :
\begin{equation}
z=ke^{(\beta-a_{2}\alpha)\chi}-\frac{\gamma}{\beta-a_{2}\alpha}.
\end{equation}
Therefore,
\begin{equation}
v=\int\frac{1}{ke^{(\beta-a_{2}\alpha)\chi}-\frac{\gamma}{\beta-a_{2}\alpha}}d\chi
,
\end{equation}
and then,
\begin{equation}
u=\frac{-1}{\gamma}log(1+\frac{\gamma}{k(\beta-a_{2}\alpha)e^{(\beta-a_{2}\alpha)
(x-\frac{y}{a_{2}})}-\gamma})+cte;
\end{equation}
where  $k=$ is arbitrary constant.\\

\noindent{}\underline{\it{Case 3.2}} : If $a_{1} =0$.\\

\noindent{In} this situation we takes $v$ as :
\begin{equation}
v=\frac{\partial}{\partial y},
\end{equation}
and the  invariants are :
\begin{eqnarray}
\chi&=&x\\
\varsigma&=&u
\end{eqnarray}
Then  $u=\varsigma$, and therefore,
\begin{eqnarray}
u_{x}&=&\varsigma_{\chi},\\
u_{y}&=&0,\\
u_{xy}&=&0.
\end{eqnarray}
Substituting these expressions into the Thomas equation, then the
solution of the corresponding equation is a constant :
\begin{equation}
\alpha u_{x} = 0, \Longrightarrow u = cte .
\end{equation}
\section {Conclusion}
\hspace{.1in} In this paper by using the criterion of invariance
of the equation under the infinitesimal prolonged infinitesimal
generators, we find the most Lie point symmetries group of the
Thomas equation. Looking the adjoint representation of the
obtained symmetry group on its Lie algebra, we have find the
preliminary classification of group-invariants solutions. We have
seen that the obtained reduced equation in such case can be
transformed on known equation by using an appropriate change of
the variables. It is intersecting to extended this construction to
the supersymmetric Thomas equation, more detail in this question
will be given elsewhere${\cite{13}}$. \newpage
\section {Appendix}
In this Appendix, we introduce the method for finding the
invariants of the group generated by the following infinitesimal
generator :

\begin{equation}
v=(a_{1}-\gamma x)\frac{\partial}{\partial x}+(a_{2}+\gamma
y)\frac{\partial}{\partial y}+(\beta x-\alpha
y)\frac{\partial}{\partial u}.
\end{equation}
The invariants are found by integrating the corresponding
characteristic system, which is
\begin{equation}
\frac{dx}{a_{1}-\gamma x}=\frac{dy}{a_{2}+\gamma
y}=\frac{du}{\beta x-\alpha y}.
\end{equation}
The first of these two equations :

\begin{equation}
\frac{dx}{a_{1}-\gamma x}=\frac{dy}{a_{2}+\gamma y},
\end{equation}

\noindent{is} easily solved, the solution are :
\begin{equation}
\frac{1}{\gamma}\log (a_{1}- \gamma x)=\frac{1}{\gamma}\log
(a_{2}+\gamma y)+c.
\end{equation}
So, one of the invariant is :

\begin{equation}
\chi =(a_{1}-\gamma x)(a_{2}+\gamma y).
\end{equation}
Note that $\chi$ is a constant for all solution of the
characteristic system, so we can replace $y$ by
$\frac{\chi}{\gamma(a_{1}-\gamma x)}-\frac{a_{2}}{\gamma}$ before
integrating. This leads to the equation :
\begin{equation}
\frac{dx}{a_{1}-\gamma x}=\frac{du}{\beta x-\frac{\alpha
\chi}{\gamma(a_{1}-\gamma x)}+\frac{\alpha a_{2}}{\gamma}};
\end{equation}
therefore :
\begin{equation}
\{\frac{\beta x}{a_{1}-\gamma
x}-\frac{\alpha\chi}{\gamma(a_{1}-\gamma x)^{2}}+\frac{\alpha
a_{2}}{\gamma(a_{1}-\gamma x)}\}dx=du.
\end{equation}
Then the equation has  the following solution : \\
\begin{equation}
\frac{-\beta}{\gamma}x-\frac{(\beta a_{1}+\alpha
a_{2})}{\gamma^{2}}\log(a_{1}-\gamma
x)-\frac{\alpha\chi}{\gamma^{2}(a_{1}-\gamma x)}=u+k,
\end{equation}
for $k$ an arbitrary constant. Finally we have :
\begin{equation}
\varsigma =u+\frac{\beta}{\gamma}x+\frac{(\beta a_{1}+\alpha
a_{2})}{\gamma^{2}}\log(a_{1}-\gamma x)+\frac{\alpha
\chi}{\gamma^{2}(a_{1}-\gamma x)},
\end{equation}
is the second invariant.

\end{document}